# In-flight calibration of NOAA POES proton detectors—Derivation of the MEPED correction factors

Marit Irene Sandanger[1], Linn-Kristine Glesnes Ødegaard[1], Hilde Nesse Tyssøy[1], Johan Stadsnes[1], Finn Søraas[1], Kjellmar Oksavik[1,2], and Kjell Aarsnes[1]

[1]Birkeland Centre for Space Science, Department of Physics and Technology, University of Bergen, Bergen, Norway, [2]Department of Arctic Geophysics, University Centre in Svalbard, Longyearbyen, Norway

**Abstract** The MEPED instruments on board the NOAA POES and MetOp satellites have been continuously measuring energetic particles in the magnetosphere since 1978. However, degradation of the proton detectors over time leads to an increase in the energy thresholds of the instrument and imposes great challenges to studies of long-term variability in the near-Earth space environment as well as a general quantification of the proton fluxes. By comparing monthly mean accumulated integral flux from a new and an old satellite at the same magnetic local time (MLT) and time period, we estimate the change in energy thresholds. The first 12 monthly energy spectra of the new satellite are used as a reference, and the derived monthly correction factors over a year for an old satellite show a small spread, indicating a robust calibration procedure. The method enables us to determine for the first time the correction factors also for the highest-energy channels of the proton detector. In addition, we make use of the newest satellite in orbit (MetOp-01) to find correction factors for 2013 for the NOAA 17 and MetOp-02 satellites. Without taking into account the level of degradation, the proton data from one satellite cannot be used quantitatively for more than 2 to 3 years after launch. As the electron detectors are vulnerable to contamination from energetic protons, the corrected proton measurements will be of value for electron flux measurements too. Thus, the correction factors ensure the correctness of both the proton and electron measurements.

## 1. Introduction

The NOAA (National Oceanic and Atmospheric Administration) POES (Polar Orbiting Environment Satellites), and MetOp (Meteorological Operational) satellites have continuously been orbiting the Earth since 1978. During this time period, 14 satellites with nearly the same instrumentation have been used to monitor the near-Earth space particle environment as well as predict the weather and climate pattern. Each satellite has a nominal lifespan of 3 years, but most satellites have been in operation much longer. NOAA 15, which was launched in 1998, is the oldest spacecraft in operation.

The protons in the inner radiation belt were first detected at low altitude by the second Soviet satellite [*Vernov et al.*, 1962]. Since then, the space age has provided us important in situ measurements of the near-Earth space environment including the ring current as well as the radiation belts. The extensive POES and MetOp series, covering more than three solar cycles (as shown in Figure 1), are currently providing the longest running estimates of the particle flux being deposited into the upper and middle atmosphere, where it can affect the chemical composition [*Jackman et al.*, 2001; *Seppälä et al.*, 2006]. However, the proton detectors on board POES and MetOp will degrade with time due to radiation damage [*Galand and Evans*, 2000] making them less sensitive to energetic particles.

It has been known for some time that solid state detectors may degrade due to radiation damage [*Coleman et al.*, 1968]. *Lyons and Evans* [1984] were the first to report nonphysical features in the MEPED (Medium Energy Proton and Electron Detector) data set that was attributed to degradation. In areas where the pitch angle distribution was expected to be isotropic or near isotropic, the proton intensities at small pitch angles often exceeded those near 90°, which is highly unlikely for energies above 30 keV. *Galand and Evans* [2000] studied the degradation of the MEPED proton detectors, and they suggested not using the data after 2–3 years for quantitative studies due to reduced sensitivity. Even though their report left no doubt that the proton detectors degrade with time, there were no immediate follow up studies. It is a complex and time-consuming task to determine the exact amount of degradation during the lifetime of each MEPED detector. It took a decade before the first study, which derived correction factors for the degradation, was done by *Asikainen and Mursula* [2011] and followed up by *Asikainen et al.* [2012] and *Ødegaard* [2013]. The studies by *Asikainen and Mursula* [2011] and *Asikainen et al.* [2012] derived correction factors for the three lowest energy channels, while *Ødegaard* [2013] derived correction factors for the two lowest energy channels. In this study, correction factors for all five energy channels are provided.

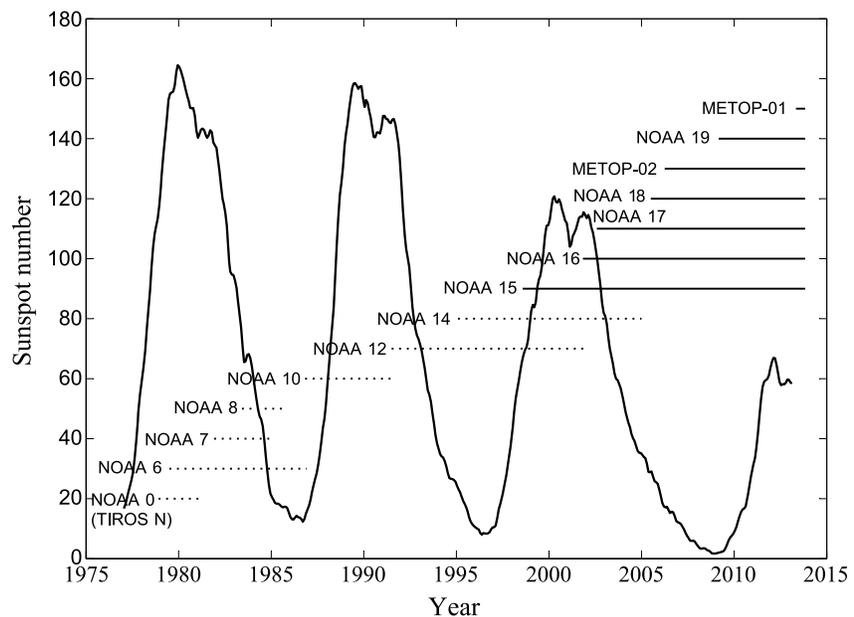

**Figure 1.** NOAA POES and MetOp satellites cover more than three solar cycles. Satellites with the SEM-1 and SEM-2 instrument package are displayed using dotted and solid lines, respectively.

In order to move from qualitative studies [e.g., *Codrescu et al.*, 1997; *Oksavik et al.*, 2000; *Søraas et al.*, 2003; *Sandanger et al.*, 2007; *Miyoshi and Kataoka*, 2008; *Horne et al.*, 2009; *Sandanger et al.*, 2009; *Rodger et al.*, 2010] to quantitative studies of protons and their effect on the environment, it is of major importance to know the MEPED detector degradation for each satellite.

The MEPED electron detectors are protected from protons below 135 keV, but higher energy protons will be able to contaminate the electron measurements. Energetic protons above 200 keV are efficiently detected by the electron detector. This is known and well understood [*Evans and Greer*, 2000; *Yando et al.*, 2011]. The contribution of protons to the electron detector response can be determined from the corrected proton observations. In order to use the MEPED electron data in quantitative studies, one has to know the flux of the energetic protons [*Evans and Greer*, 2000; *Yando et al.*, 2011; *Whittaker et al.*, 2014]. The MEPED proton correction factors are thus also needed for the quantitative use of the electron data. The main purpose of this paper is to enable the use of proton and electron data from POES MEPED detectors even after the proton detectors start to degrade.

We start in section 2 with an orbital overview of the POES and MetOp satellites where we focus on the satellites NOAA 15 and above, with the SEM-2, a newer and modified SEM (Space Environment Monitor) instrument package, followed by section 3 describing the degradation and radiation damage of the MEPED proton detectors. Section 4 explains our method for deriving the correction factors. We use all available data throughout the whole orbit. Only satellites in the same MLT are compared. The monthly mean integral flux spectrum from an old satellite is compared with that from a new satellite for the same month. By comparing monthly mean spectra, it is expected that rapid fluctuations in the data are leveled out and that the two satellites are subjected to the same average particle environment. The 12 monthly energy spectra from the first year of a new satellite are used as a reference. Section 5 gives an overview of the correction factors, while section 6 explains how to obtain the monthly correction factors for times when direct calibration is not available. We show how the temporal evolution of our correction factors compares with results from earlier studies [*Asikainen and Mursula*, 2011; *Asikainen et al.*, 2012]. In section 7 the corrected proton flux at fixed energies from satellites of different operation times are compared. The fluxes overlap close to perfectly, verifying of the quality of the derived correction factors. We also demonstrate examples of uncorrected and corrected data for a ten year old satellite to visualize the importance of correcting the data. Finally, we discuss reliability and limitation of our method and results.

Table 1. Overview of the Different Energy Channels (Differential and Integral) of the SEM-2 MEPED Proton Detectors for Both the 0° and the 90° Detector[a]

| Channel Identification | Proton Energy (keV) | |
|---|---|---|
| | Differential Channel | Integral Channel |
| 0°/90° P1 | 30 to 80 | >30 |
| 0°/90° P2 | 80 to 250 | >80 |
| 0°/90° P3 | 250 to 800 | >250 |
| 0°/90° P4 | 800 to 2500 | >800 |
| 0°/90° P5 | 2500 to 6900 | >2500 |
| 0°/90° P6 | >6900 | |

[a]The flux in the integral channels is derived from the differential channels. The flux in the P6 differential channel is not used to derive integral flux, due to possible contamination by relativistic electrons [*Yando et al.*, 2011].

## 2. NOAA POES and MetOp Satellites

The NOAA POES and MetOp satellites are Sun-synchronous low-altitude polar orbiting spacecraft. Their orbital period is about 103 min, resulting in 14–15 orbits each day. The NOAA POES and the MetOp satellites together cover more than three solar cycles, with the first spacecraft NOAA 0 (TIROS-N) launched in 1978. The satellites from NOAA 0 up to NOAA 14 carried the first version of the instrument package, SEM-1, which varies slightly in energy bands from the SEM-2. In the current paper we focus on the newer SEM-2 instrument package used on NOAA 15 launched in 1998, and up until MetOp-01 launched in late 2012. The satellite MetOp-03 is planned to be launched in 2017 and will be the last one in this series carrying the instrument package SEM-2.

Figure 1 displays the operational period of the satellites with the SEM-1 (dotted lines) and SEM-2 (solid lines) instrument packages. The SEM MEPED instrument consists of two identical proton detectors, one viewing nearly radially outward from Earth and the other viewing nearly antiparallel to the satellite's velocity vector; for details, see *Evans and Greer* [2000]. These two detectors will be referred to as the 0° and the 90° detectors. At high latitudes this is approximately the pitch angle of the particles being measured by the respective detectors. The detectors have an opening angle of 30° full width half maximum. Table 1 gives an overview of the nominal energy thresholds of the SEM-2 MEPED proton detectors. The P6 channel (>6900 keV) sometimes experiences contamination from relativistic electrons. Except for these periods, this channel has very low count rates compared to the other channels. Based on this fact, we have dropped the P6 channel from our analysis. We then treat the P5 channel (2500–6900 keV) as an integral channel for >2500 keV protons.

The satellite orbits are located at different magnetic local time sectors, as visualized in Figure 2. The orbital planes show a variable drift in local time, and as shown in Figure 2, NOAA 15 and NOAA 16 have an especially large drift in local time. This drift has been previously shown by *Asikainen et al.* [2012]. Table 2 gives an overview of the satellites' operation time, mean altitude, ascending node at launch, and which satellites are intercalibrated. The satellites are launched alternately in low-altitude (∼825 km) morning orbits, and high-altitude (∼865 km) afternoon orbits. In the calibration we are comparing satellites which have the same local time orbits and, therefore, due to the launch procedure, also fly in the same orbit height. NOAA 16 is an exception, as the orbit is drifting into new local time sectors during its lifetime. Due to possible biases in several of its monthly spectra in 2013 we do not compare it with MetOp-01.

The MEPED instrument measures both protons and electrons, but we will focus on the proton detector which is more affected by radiation damage compared to the electron detector, as the electron detector has a nickel foil protecting it from protons below 135 keV [*Galand and Evans*, 2000]. The proton detectors are equipped with broom magnets excluding electrons with energies below 1.5 MeV to be detected. MEPED cannot distinguish between different ion species, but we will use the term protons to represent all ions measured by MEPED. A full description of the SEM-2 instrument package is given by *Evans and Greer* [2000], while the older SEM-1 instrument package is described by *Raben et al.* [1995].

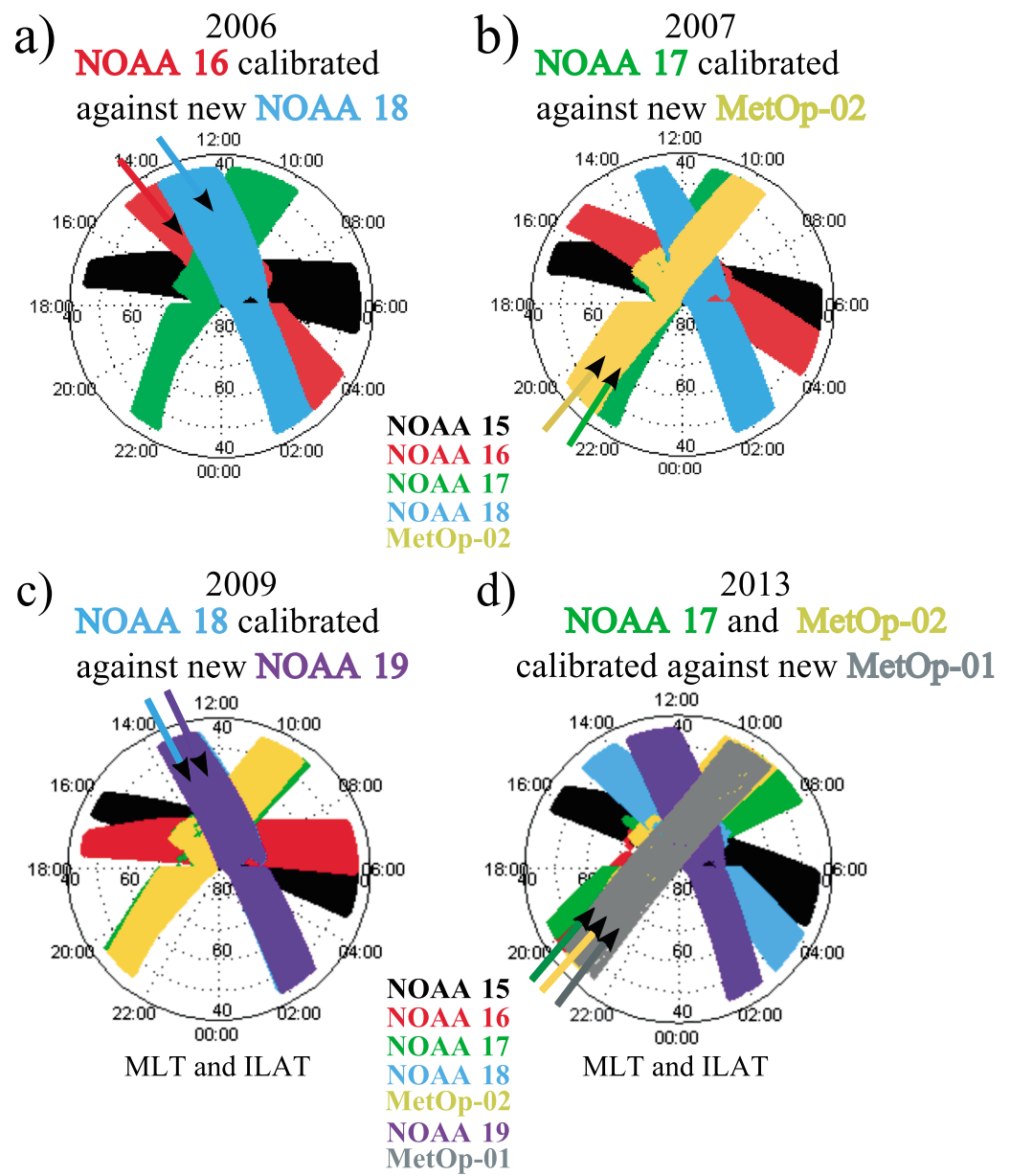

**Figure 2.** Footprints of the NOAA/POES and MetOp spacecrafts given as invariant latitude and magnetic local time in the Northern Hemisphere. (a) February 2006 when NOAA 18 is newly launched and have overlapping footprints with NOAA 16. (b) July 2007 when MetOp-02 is newly launched and calibrates NOAA 17. (c) September 2009 when NOAA 19 is newly launched and calibrates NOAA 18. (d) Data from 2013 when the footprint of NOAA 16, NOAA 17, and MetOp-02 are underneath the footprint of MetOp-01. MetOp-01 calibrates only NOAA 17 and MetOp-02 due to possible biases in several of the NOAA 16 monthly spectra in 2013.

## 3. Radiation Damage

Protons of energies between 50 keV and 5 MeV can generate significant radiation damage in silicon surface-barrier detectors [*Coleman et al.*, 1968]. The MEPED instrument measures protons in this energy range. *Galand and Evans* [2000] showed that the amount of proton flux measured by the MEPED detector is large enough to induce serious damage to the instrument.

**Table 2.** List of NOAA POES and MetOp Satellites Operation Time, Mean Altitude, Ascending Nodes at Launch, and Which Satellites They Calibrate[a]

| Spacecraft | Operation | Altitude | Ascending Node | Calibrates | Figure 2 Panel |
|---|---|---|---|---|---|
| NOAA 15 | 1998–present | 821 | 19:30 | | |
| NOAA 16 | 2001–2014 | 862 | 14:00 | | |
| NOAA 17 | 2002–2013 | 823 | 22:00 | | |
| NOAA 18 | 2005–present | 866 | 14:00 | NOAA 16 | (a) |
| MetOp-02 | 2006–present | 832 | 21:30 | NOAA 17 | (b) |
| NOAA 19 | 2009–present | 868 | 14:00 | NOAA 18 | (c) |
| MetOp-01 | 2012–present | 832 | 21:30 | NOAA 17 | (d) |
| MetOp-01 | 2012–present | 832 | 21:30 | MetOp-02 | (d) |

[a]The given mean altitude is for the year 2009. The ascending node is the local time when the spacecraft cross the equator in a northerly direction. Last column refers to the panel in Figure 2 that visualizes the satellites overlapping orbits in MLT which gives grounds for calibration. Section 4 will focus on the last column in the table (which shows the satellite that can be calibrated).

The NOAA Technical Report [*Galand and Evans*, 2000] gives a thorough evaluation of the MEPED proton instrument's radiation damage. The MEPED electron detector, on the other hand, did not suffer as badly from radiation damage. This can be explained by the electron detector's nickel foil, which prevents any protons of energy below 135 keV from reaching the detector and also considerably reduces the energy flux of protons of higher energies [*Galand and Evans*, 2000; *Yando et al.*, 2011].

During its lifetime, the 90° detector accumulates more counts than the 0° detector due to the pitch angle distribution (PAD) of the particles. One therefore expects that the 90° detector deteriorates faster than the 0° detector [*Galand and Evans*, 2000]. To determine the rate of degradation, *Galand and Evans* [2000] examined the ratio of the 0° response versus 90° response at geomagnetic latitudes above 60°. This ratio reflects the level of anisotropy of the radiation. Since anisotropic conditions during an orbit are more common than isotropic conditions, this ratio is usually low. Nevertheless, *Galand and Evans* [2000] found that the number of cases with the ratio >1 increased with time for all four NOAA satellites investigated (NOAA 0, NOAA 6, NOAA 10, and NOAA 12), confirming that the 90° proton detector degrades faster than the 0° detector, consistent with a pancake-shaped PAD of the protons. This comparison was performed using all SEM-1 instruments and for different levels of solar activity.

*Galand and Evans* [2000] plotted the occurrence rate of higher responses in the 0° detector than in the 90° detector, and they found that the slope and shape of the increase varied from one satellite to the next, implying different degradation rates. For NOAA 6, launched during solar maximum, the occurrence rate reached higher values for a much smaller total accumulated counts compared to NOAA 10, which was launched during solar minimum. They suggested that the radiation damage may not be a linear function of the total number of accumulated counts; the count rate may also play a role. It is also possible that the protons measured by the MEPED are not the only particles responsible for the radiation damage. Protons of lower energies, as well as electrons, relativistic protons with energies >100 MeV (from solar proton events), and heavy ions can also participate in the deterioration of the instruments.

*Galand and Evans* [2000] highlighted two types of possible damage:

1. Dead layer. The formation of a dead layer in which the incident particle is slowed down and where part of its energy is absorbed without contributing to the charge collection (not measured).
2. Partial charge collection. A decrease in the mobility of the free electrons/holes created in the crystal by the incident proton. The free electrons/holes recombine before they have contributed fully to the charge collection.

Due to the two effects, the charge collection is reduced and the energy of the incident proton will thereby be underestimated. As the detector degrades, the energy of the protons must increase in order to trigger a fixed energy threshold, and a new and an old detector will be sensitive to different parts of the particle population. A new detector measures protons with energies as low as 30 keV (see Table 1), whereas the most degraded detector in our study (see NOAA 17 in Table 4), cannot detect protons with energy lower than 60 keV.

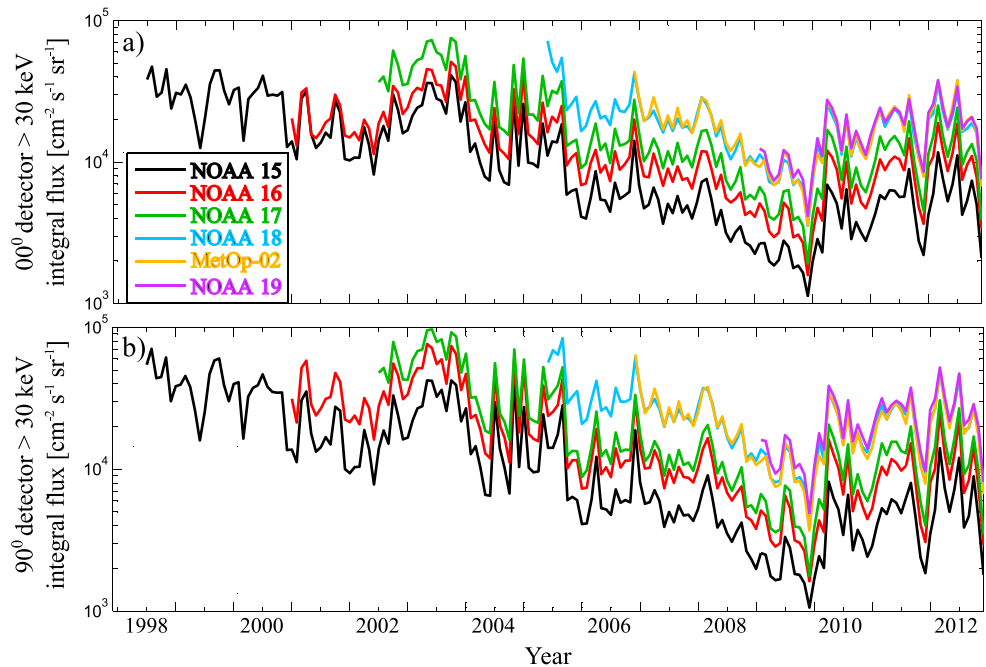

**Figure 3.** The monthly accumulated proton integral flux for the satellites NOAA 15, 16, 17, 18, and 19 and MetOp-02 with nominal energy greater than 30 keV. (a) Data from the 0° detector. (b) Data from the 90° detector.

Figure 3 shows the monthly mean proton integral flux (for the uncorrected nominal energy threshold E > 30 keV) from 1998 to the end of 2013 for the NOAA 15, NOAA 16, NOAA 17, NOAA 18, MetOp-02, and NOAA 19 satellites. The data from the 0° detector are displayed in Figure 3a, and data from the 90° detector is displayed in Figure 3b. Satellites in similar orbits with identical instruments should measure the same particle flux variations when monthly averages are compared. Figure 3 clearly shows indications of detector degradation, as an old satellite measures lower fluxes than a new satellite in the same time period at nominal energies. The discrepancy between the flux measured by an old and a new satellite can be nearly an order of magnitude. It is worth noticing how the 90° detector (Figure 3b) shows higher fluxes than the 0° detector (Figure 3a), even though the 90° detector suffers more damage than the 0° detector. The 90° detector measures up to 50% more than the 0° detector and on an average 25% more, due to the pancake shaped PAD.

## 4. Method to Determine the Correction Factors

We want to estimate the current energy thresholds of the MEPED proton detector. The In-Flight-Calibration files for all the satellites have been examined, and the noise level in all of them, in the time intervals when the $\alpha$ factors have been estimated, is nominal. The ratio of the current energy thresholds to the nominal energy thresholds is called the correction factor, $\alpha$. The $\alpha$ factors give the energy changes of the degraded detector. In that way we can establish corrected integral spectra. To achieve this, we compare particle spectra from a degraded detector with a new and nondegraded detector.

A magnetic storm will increase the proton fluxes more in the MLT evening and nightside sector than in the morning and noon sector [*Codrescu et al.*, 1997], on timescales ranging from a few hours to a few days. Due to such magnetic local time effects, an old satellite is only compared to a satellite in the same MLT sector. NOAA 15, NOAA 17, MetOp-02, and MetOp-01 have ascending nodes in the evening MLT region, while NOAA 16, NOAA 18, and NOAA 19 have ascending nodes in the afternoon sector. It is important to account for this systematic behavior, and therefore, we only compare a new satellite with an old satellite at the same MLT. The NOAA 15 and NOAA 16 drift in their orbits, making them hard to compare with newer satellites (see last column of Table 2 for an overview of comparable satellites).

In our method the data from all latitudes and longitudes are used. For each satellite, the monthly mean of the proton flux is calculated for each energy channel. If data for a whole day is missing for one satellite, then the same day is removed from all comparable satellite's data sets in order for the satellites to measure in the same time intervals. The monthly mean differential flux is then converted to a monthly mean integral flux. Figure 4 illustrates how the $\alpha$ factors are derived based on the monthly mean accumulated integral flux energy spectra from a new satellite (NOAA 18) and a 4.5 year old satellite (NOAA 16). Henceforth, the $\alpha$ factors derived from this calibration method will be referred to as the $\alpha_{cal}$ factor.

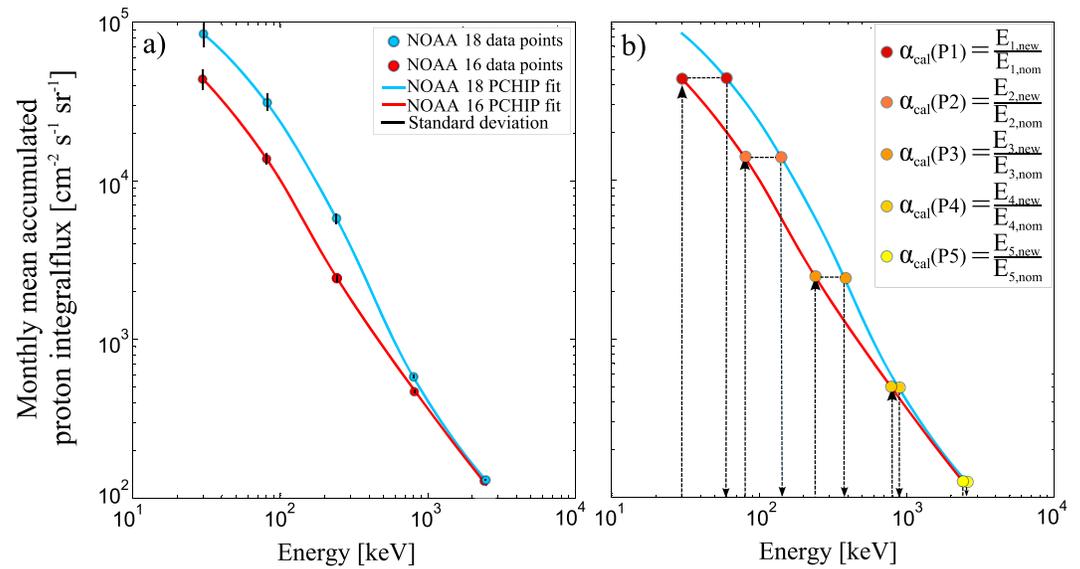

**Figure 4.** Illustration of how the $\alpha_{cal}$ factors are derived. The monthly mean accumulated proton energy spectra from a new (NOAA 18) and 4.5 years old (NOAA 16) satellite are shown in the panels. The data are from the 90° detectors during October 2005. (a) The integral flux from NOAA 18 and from NOAA 16 with a PCHIP fit to the data plotted on a logarithmic scale. (b) The procedure of determining the new energy thresholds of the older NOAA 16 satellite from the NOAA 18 energy spectrum. The standard deviations are shown as error bars in panel (a).

We have chosen to use the mean and include zero count rate values within this calculation. This inclusion is due to low fluxes, which might not trigger a count in the instrument. It should be noted that there is a potential for error by including zero counts. Hence, our use of the mean is essential with a large number of zero values and again this could be a source of error in a skewed distribution. The final results are discussed in section 7, which show a good correlation between an old satellite with an alpha correction and a new satellite and indicate that this method of analysis is justified and the errors are assumed to be minimal.

In order to numerically represent the integral spectrum, we have fitted a monotonic Piecewise Cubic Hermite Interpolating Polynomial (PCHIP) to the measurements, illustrated in Figure 4a. A PCHIP fit is guaranteed to go through all your data points, as well as producing a monotone function that is physically consistent with an integral spectrum. Figure 4b demonstrates the method and the basic idea behind correcting for the degradation of a detector. The new detector measures the observed energy spectrum at the nominal energy thresholds, while the degraded detector measures the same energy spectrum, with the important difference that the energy thresholds are unknown. With the assumption that the old satellite's integral flux level is correct, and the energy threshold is the unknown, the procedure is as follows: The integral flux at the nominal energy threshold for the old satellite is found (follow the arrows in Figure 4b), and for that specific integral flux the corresponding energy threshold of the new satellite is found. This is done for all five nominal energy thresholds.

The respective flux and the associated energy give us the increased energy threshold for the particles detected:

$$\alpha_{cal} = \frac{E_{new}}{E_{nom}} \quad (1)$$

Each comparison between the energy spectrum from a new and an old satellite generates a set of $\alpha_{cal}$ factors

$$\text{Correction factors} = [\alpha_{cal}(P1), \alpha_{cal}(P2), \alpha_{cal}(P3), \alpha_{cal}(P4), \alpha_{cal}(P5)] \quad (2)$$

that corresponds to the five integral channels P1–P5 (as shown in Figure 4b). The nominal energy of the integral channels P1–P5 is given in the last column in Table 1.

**Table 3.** The $\alpha_{cal}$ Factors for All Four Satellites and for Both the 0° and 90° Detectors[a]

| Satellite | Year (Month) | Mean $\alpha_{cal}$ (Standard Deviation) | | | | |
|---|---|---|---|---|---|---|
| | | $\alpha_1$ | $\alpha_2$ | $\alpha_3$ | $\alpha_4$ | $\alpha_5$ |
| | | 0° Detector | | | | |
| NOAA 16 | 2005 (Feb) | 1.57 (0.08) | 1.65 (0.08) | 1.22 (0.05) | 1.12 (0.06) | 1.08 (0.08) |
| NOAA 17 | 2007 (Jul) | 1.37 (0.06) | 1.59 (0.07) | 1.19 (0.02) | 1.07 (0.01) | 1.03 (0.01) |
| NOAA 17 | 2013 (Apr) | 1.53 (0.05) | 1.82 (0.10) | 1.32 (0.07) | 1.27 (0.09) | 1.16 (0.08) |
| NOAA 18 | 2009 (Sep) | 1.06 (0.04) | 1.14 (0.05) | 1.07 (0.01) | 1.16 (0.01) | 1.17 (0.01) |
| MetOp-02 | 2013 (Apr) | 1.13 (0.07) | 1.27 (0.10) | 1.09 (0.07) | 1.04 (0.06) | 1.05 (0.06) |
| | | 90° Detector | | | | |
| NOAA 16 | 2005 (Feb) | 1.89 (0.10) | 1.86 (0.07) | 1.31 (0.02) | 1.08 (0.03) | 1.03 (0.03) |
| NOAA 17 | 2007 (Jul) | 1.65 (0.08) | 1.94 (0.07) | 1.46 (0.02) | 1.25 (0.01) | 1.21 (0.01) |
| NOAA 17 | 2013 (Apr) | 1.94 (0.06) | 2.10 (0.10) | 1.39 (0.09) | 1.22 (0.06) | 1.21 (0.05) |
| NOAA 18 | 2009 (Sep) | 1.11 (0.05) | 1.28 (0.06) | 1.12 (0.01) | 1.05 (0.01) | 1.03 (0.01) |
| MetOp-02 | 2013 (Apr) | 1.30 (0.09) | 1.47 (0.10) | 1.10 (0.06) | 0.92 (0.05) | 0.92 (0.05) |

[a]Columns 3–7 show the mean $\alpha_{cal}$ factors with the standard deviation value in parentheses.

The $\alpha_{cal}$ factors are derived for each month throughout the first year after launch of a new satellite in the same MLT range as that of the old satellites. The sets of 12 monthly $\alpha_{cal}$ factors are used to determine the mean $\alpha_{cal}$ factors for the respective years, and the standard deviation is used as a measure of the spread in the data.

## 5. Resulting $\alpha_{cal}$ Factors

With our method we retrieve two sets of $\alpha_{cal}$ factors for NOAA 17 at the age of 4 and 10 years. For NOAA 16, NOAA 18, and MetOp-02 we retrieve only one set of $\alpha_{cal}$ factors, the NOAA 16 and 18 at the age of 4 and for MetOp-02 at 6 years. NOAA 15, NOAA 19, and MetOp-01 are without any $\alpha_{cal}$ factor at present time. In 2016/2017, the MetOp-03 satellite will be launched. This new MetOp satellite will orbit the Earth in the same MLT range as the MetOp-01 and 02 satellites, giving us the opportunity to derive $\alpha_{cal}$ factors for NOAA 17, MetOp-02, and MetOp-01 (all in the same MLT sector).

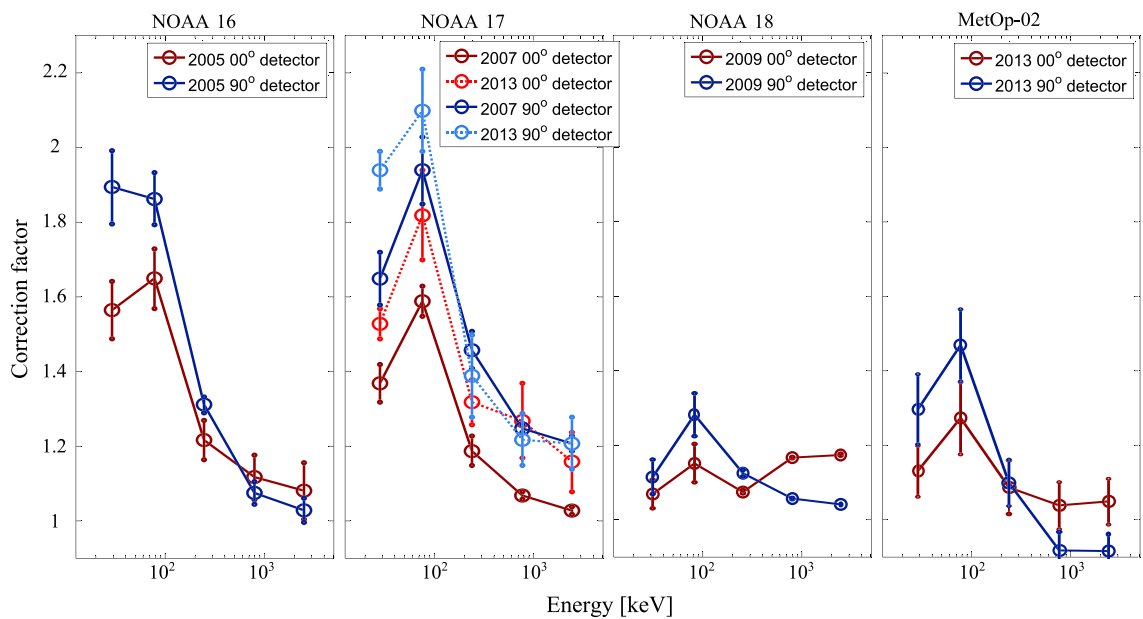

**Figure 5.** The different $\alpha_{cal}$ factors as a function of energy for NOAA 16, NOAA 17, NOAA 18, and MetOp-02. The blue lines display the $\alpha_{cal}$ factors for the 90° detector, while the red lines display the 0° detector. In the second panel, the dotted lines display the $\alpha_{cal}$ factors for 2013. The mean standard deviation is marked as a vertical error bar in each point.

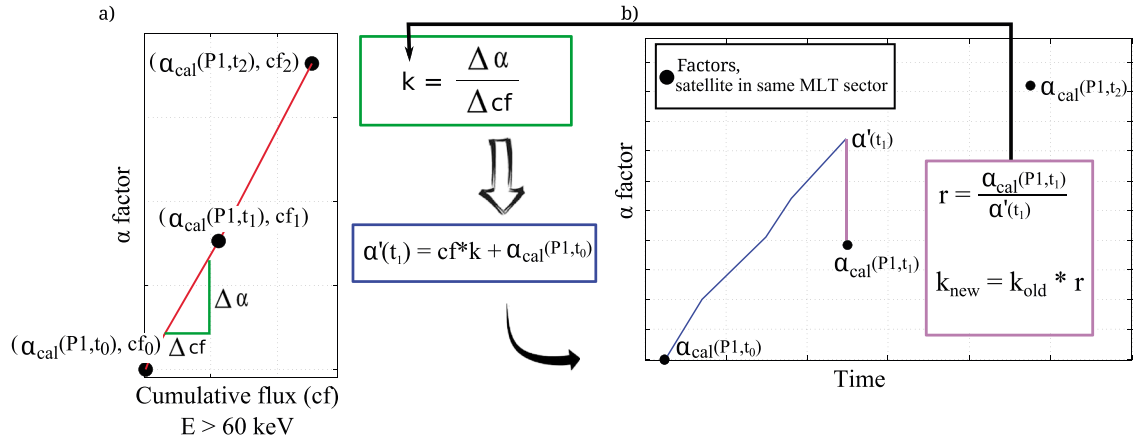

**Figure 6.** A visualization of the iteration process that is used to find the monthly $\alpha_{month}$ factors in between two $\alpha_{cal}$ factors. (a) The $\alpha_{cal}$ factors versus the calculated accumulated integral flux at the time when we have determined the $\alpha_{cal}$ factors. The linear slope $k$ is calculated and used in the calculation of an $\alpha'$ describing the $\alpha$ factor after the first month. (b) The $\alpha$ factors versus time, showing the same three $\alpha_{cal}$ points as in Figure 6a, but also the new point $\alpha'$. Figure 6b shows how we need an iterative process to find the right constant $k$ in the linear equation.

Our mean $\alpha_{cal}$ factors together with the associated standard deviation values are shown in Table 3.

Figure 5 displays the $\alpha_{cal}$ factors versus energy for the four NOAA POES and MetOp satellites. It illustrates how the 90° detector (blue color scale) degrades faster than the 0° detector (red color scale). It is also evident from Figure 5 and Table 3 that the lowest energy channel (P1) degrades slower than the second energy channel (P2). This feature is somewhat unexpected and is commented further in relation to other studies at the end of section 6.1.

## 6. Temporal Evolution of the Degradation

The $\alpha$ factor is calculated at points separated by several years, and it is thus important to find a method of interpolation in between the calibration points. In this section we show a procedure for finding the monthly correction factors, $\alpha_{month}$, throughout a satellites's lifetime. Assuming that it is the number of particles that hit the detector that is the main cause of the degradation, we use the method of accumulative flux to achieve the temporal evolution of the degradation.

To get an estimation of $\alpha_{month}$ based on the flux in the period between two $\alpha_{cal}$ factors, we go through an iterative process illustrated in Figure 6. This process is done for both detectors and all channels separately.

The first step is to apply the uncorrected integral flux, accumulated from the satellite was new and until the time of our $\alpha_{cal}$ factors. The uncorrected accumulative integral flux at these points is denoted $cf_t$, where the time is given by $t$. The $\alpha_{cal}$ factor for energy channel P1 is denoted $\alpha_{cal}(P1)$, and for time $t = 0$ the factor is denoted $\alpha_{cal}(P1, t_0)$. The $\alpha_{cal}$ factor and $cf$ at $t = 0$, when the satellite is new, is respectively 1 and 0 for all energy channels. In Figure 6a $\{\alpha_{cal}(P1, t_0), cf_0\}$, $\{\alpha_{cal}(P1, t_1), cf_1\}$ and $\{\alpha_{cal}(P1, t_2), cf_2\}$ are plotted as black points.

The second step of the process is to calculate the linear slope $k$ between the two first points in Figure 6a.

$$k = \frac{\Delta \alpha}{\Delta cf} = \frac{\alpha_{cal}(P1, t_1) - \alpha_{cal}(P1, t_0)}{cf_1 - cf_0} \tag{3}$$

The third step is to find the accumulated flux 1 month after the launch of the satellite. The flux is not accumulated $> E_{nom}$ thresholds, but rather the $> E_{new}$ thresholds. In that way, we ensure to accumulate the correct particle population. Lower energy protons can still degrade the detectors, but the $E_{new}$ thresholds are set at the energies where even the most degraded satellite is still able to measure:

$$E_{new} = [60, 167, 360, 1050, 2300] \text{ (keV)} \tag{4}$$

in contrary to the nominal energy thresholds:

$$E_{nom} = [30, 80, 250, 800, 2500] \text{ (keV)} \tag{5}$$

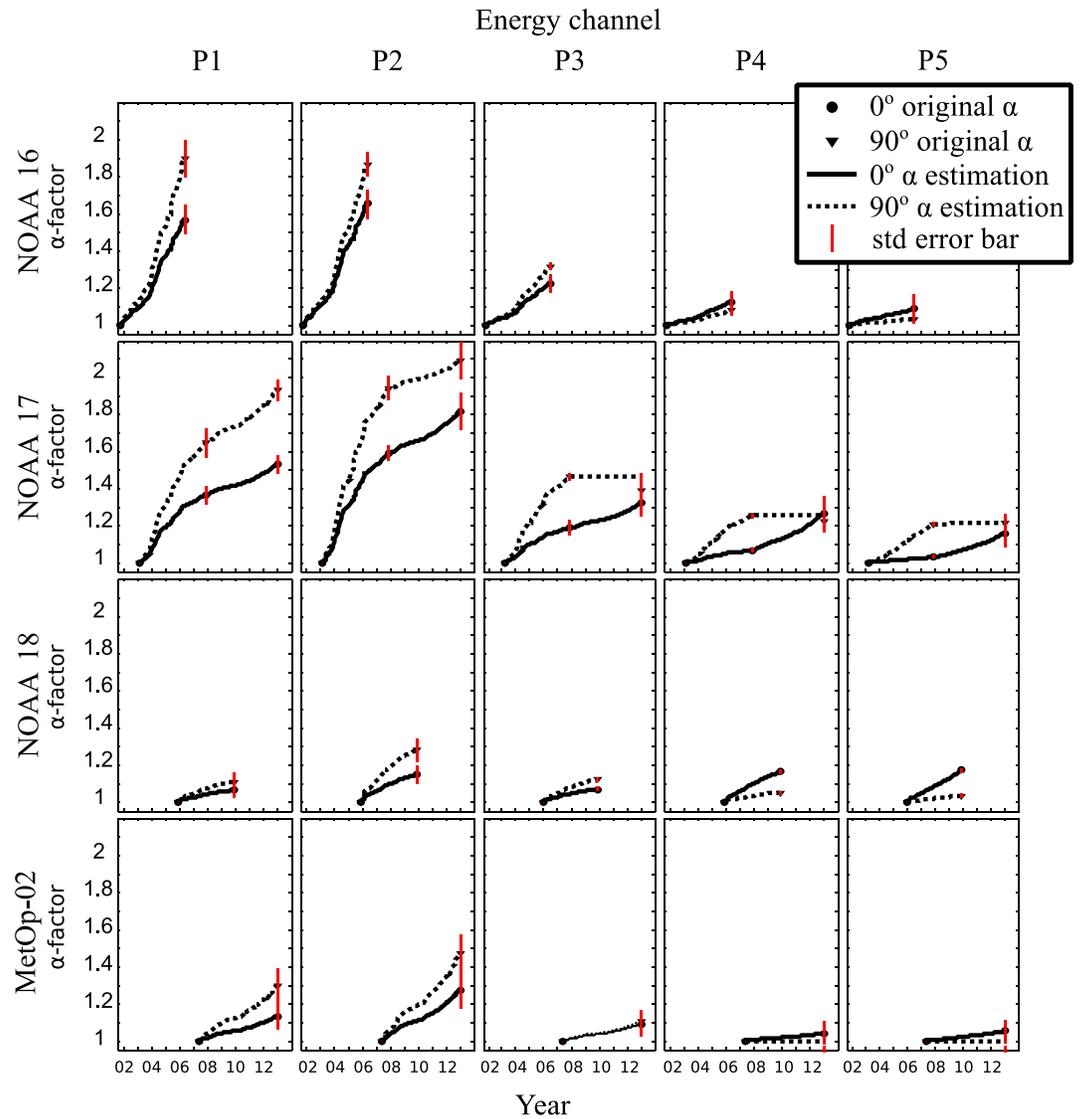

**Figure 7.** Each of the subplots display the $\alpha_{month}$ factor versus time for all four NOAA POES and MetOp satellites. The circles and triangles mark the $\alpha_{cal}$ factors from the 0° and the 90° detectors, respectively. The line in between these points is found through flux estimation, the dotted lines exhibit the 90° detector, and the solid lines exhibit the 0° detector. The MAD error bar is shown as red vertical lines.

The fourth step is to use the accumulated flux 1 month after the launch of a new satellite, together with the calculated slope $k$ found in equation (3) in order to find the $\alpha_{month}$ factor after 1 month, $\alpha'$:

$$\alpha' = cf \times k + \alpha_{cal}(P1, t_0) = cf \times k + 1 \qquad (6)$$

The calculated $\alpha'$ factor for the first month is used to correct the energy threshold and thus the accumulated flux of the second month in the satellite's lifetime, before an $\alpha'$ factor is calculated for the second month. This process is repeated until we reach the month at time $t_1$ corresponding to the month where we have our $\alpha_{cal}(P1, t_1)$. By this point, we have reached an $\alpha'(t_1)$ factor that is larger or smaller than the $\alpha_{cal}(P1, t_1)$. We find the ratio:

$$r = \frac{\alpha_{cal}(P1, t_1)}{\alpha'(t_1)} \qquad (7)$$

and adjust the slope $k$ by this factor, i.e.,

$$k_{new} = k_{old} \times r \qquad (8)$$

Table 4. The $\alpha_{year}$ Factors for NOAA 16, NOAA 17, NOAA 18, and MetOp-02 Satellites for the 0° and the 90° Detectors[a]

| Satellite | Year | 0° Detector | | | | | 90° Detector | | | | |
| --- | --- | --- | --- | --- | --- | --- | --- | --- | --- | --- | --- |
| | | $\alpha_1$ | $\alpha_2$ | $\alpha_3$ | $\alpha_4$ | $\alpha_5$ | $\alpha_1$ | $\alpha_2$ | $\alpha_3$ | $\alpha_4$ | $\alpha_5$ |
| NOAA 16 | 2001 | 1.03 | 1.04 | 1.01 | 1.01 | 1.01 | 1.05 | 1.05 | 1.02 | 1.01 | 1.00 |
| NOAA 16 | 2002 | 1.09 | 1.11 | 1.04 | 1.02 | 1.03 | 1.13 | 1.13 | 1.05 | 1.01 | 1.01 |
| NOAA 16 | 2003 | 1.21 | 1.24 | 1.08 | 1.04 | 1.04 | 1.30 | 1.29 | 1.11 | 1.03 | 1.02 |
| NOAA 16 | 2004 | 1.37 | 1.43 | 1.14 | 1.08 | 1.06 | 1.53 | 1.51 | 1.19 | 1.03 | 1.02 |
| **NOAA 16** | **2005** | **1.51** | **1.59** | **1.20** | **1.11** | **1.08** | **1.79** | **1.76** | **1.28** | **1.07** | **1.03** |
| NOAA 17 | 2002 | 1.00 | 1.00 | 1.00 | 1.00 | 1.00 | 1.00 | 1.00 | 1.00 | 1.00 | 1.00 |
| NOAA 17 | 2003 | 1.08 | 1.13 | 1.04 | 1.01 | 1.00 | 1.13 | 1.19 | 1.10 | 1.04 | 1.02 |
| NOAA 17 | 2004 | 1.19 | 1.31 | 1.10 | 1.03 | 1.01 | 1.31 | 1.44 | 1.22 | 1.11 | 1.06 |
| NOAA 17 | 2005 | 1.27 | 1.44 | 1.14 | 1.05 | 1.02 | 1.47 | 1.68 | 1.33 | 1.17 | 1.12 |
| NOAA 17 | 2006 | 1.33 | 1.52 | 1.17 | 1.06 | 1.02 | 1.57 | 1.83 | 1.41 | 1.22 | 1.16 |
| **NOAA 17** | **2007** | **1.37** | **1.59** | **1.19** | **1.07** | **1.03** | **1.64** | **1.94** | **1.46** | **1.25** | **1.21** |
| NOAA 17 | 2008 | 1.40 | 1.63 | 1.21 | 1.10 | 1.04 | 1.70 | 1.97 | 1.46 | 1.25 | 1.21 |
| NOAA 17 | 2009 | 1.41 | 1.66 | 1.22 | 1.12 | 1.06 | 1.73 | 1.99 | 1.46 | 1.25 | 1.21 |
| NOAA 17 | 2010 | 1.44 | 1.69 | 1.24 | 1.15 | 1.09 | 1.77 | 2.01 | 1.46 | 1.25 | 1.21 |
| NOAA 17 | 2011 | 1.47 | 1.73 | 1.27 | 1.19 | 1.11 | 1.83 | 2.04 | 1.46 | 1.25 | 1.21 |
| NOAA 17 | 2012 | 1.51 | 1.79 | 1.30 | 1.24 | 1.14 | 1.90 | 2.08 | 1.46 | 1.25 | 1.21 |
| **NOAA 17** | **2013** | **1.53** | **1.82** | **1.32** | **1.27** | **1.16** | **1.94** | **2.10** | **1.46** | **1.25** | **1.21** |
| NOAA 18 | 2005 | 1.00 | 1.00 | 1.00 | 1.00 | 1.00 | 1.00 | 1.00 | 1.00 | 1.00 | 1.00 |
| NOAA 18 | 2006 | 1.02 | 1.05 | 1.02 | 1.05 | 1.04 | 1.04 | 1.10 | 1.04 | 1.02 | 1.01 |
| NOAA 18 | 2007 | 1.04 | 1.09 | 1.04 | 1.09 | 1.08 | 1.07 | 1.18 | 1.08 | 1.03 | 1.02 |
| NOAA 18 | 2008 | 1.05 | 1.12 | 1.06 | 1.13 | 1.12 | 1.09 | 1.24 | 1.10 | 1.04 | 1.02 |
| **NOAA 18** | **2009** | **1.06** | **1.14** | **1.07** | **1.15** | **1.16** | **1.10** | **1.27** | **1.11** | **1.05** | **1.03** |
| MetOp-02 | 2007 | 1.01 | 1.03 | 1.01 | 1.00 | 1.00 | 1.03 | 1.05 | 1.01 | 1.00 | 1.00 |
| MetOp-02 | 2008 | 1.04 | 1.08 | 1.03 | 1.01 | 1.01 | 1.09 | 1.14 | 1.03 | 1.00 | 1.00 |
| MetOp-02 | 2009 | 1.05 | 1.10 | 1.03 | 1.02 | 1.02 | 1.12 | 1.18 | 1.04 | 1.00 | 1.00 |
| MetOp-02 | 2010 | 1.07 | 1.14 | 1.04 | 1.02 | 1.03 | 1.15 | 1.24 | 1.05 | 1.00 | 1.00 |
| MetOp-02 | 2011 | 1.09 | 1.18 | 1.06 | 1.03 | 1.04 | 1.20 | 1.31 | 1.07 | 1.00 | 1.00 |
| MetOp-02 | 2012 | 1.12 | 1.25 | 1.08 | 1.04 | 1.05 | 1.26 | 1.42 | 1.09 | 1.00 | 1.00 |
| **MetOp-02** | **2013** | **1.13** | **1.27** | **1.09** | **1.04** | **1.05** | **1.30** | **1.47** | **1.10** | **1.00** | **1.00** |

[a]The $\alpha_{year}$ factors are from the midpoint of each year. The $\alpha_{year}$ factors given in bold text are from the same years when we have $\alpha_{cal}$ factors.

The process starts again from the first month after launch with the slope $k_{new}$ as input in equation (6), that is:

$$\alpha' = cf \times k_{new} + \alpha_{cal}(P1, t_0) = cf \times k_{new} + 1 \quad (9)$$

The process is then repeated until $\alpha'(t_1) = \alpha_{cal}(P1, t_1)$. The same process is also applied between factors $\alpha_{cal}(P1, t_1)$ and $\alpha_{cal}(P1, t_2)$.

The $\alpha_{month}$ factors based on the accumulated flux method is shown with a monthly resolution in Figure 7 for NOAA 16, NOAA 17, NOAA 18, and MetOp-02.

The resulting yearly $\alpha$ factors, for NOAA 16, NOAA 17, NOAA 18, and MetOp-02 are given in Table 4. Henceforth, these yearly $\alpha$ factors will be referred to as the $\alpha_{year}$ factors and are taken from the month in the middle of each year. The $\alpha_{year}$ factors given in bold text in Table 4 are from the same years when we have $\alpha_{cal}$ factors. But the $\alpha_{year}$ factor with bold text may not be identical to the values in Table 3 since the $\alpha_{cal}$ factors in Table 3 are placed at the midpoint of the first 12 months after launch of a new satellite, while the $\alpha_{year}$ factors are from the middle of the calender year. Also, the accumulative flux method does not allow the $\alpha_{month}$ factors and $\alpha_{year}$ factors to decrease. This can lead to a constant $\alpha$ factor as shown in Figure 7 (third panel, NOAA 17). The 90° detector's $\alpha_{cal}(P3, t_2)$ value is slightly less than the $\alpha_{cal}(P3, t_1)$ value but can be kept constant within the error bars.

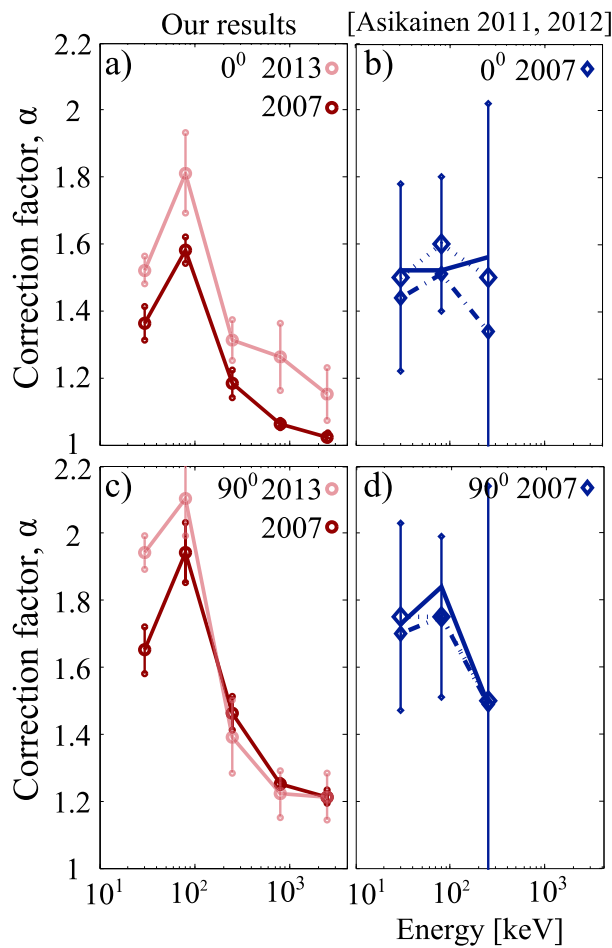

**Figure 8.** A comparison between the $\alpha_{cal}$ factors of the (a and b) 0° and (c and d) 90° detector of the two different studies. Our $\alpha$ factors for NOAA 17, year 2007 and 2013 are shown in, respectively, solid and dotted lines (Figures 8a and 8c). The $\alpha_{cal}$ factor for the same satellite in 2007 retrieved by *Asikainen and Mursula* [2011] is shown in dotted and dashed blue lines in Figures 8b and 8d, while *Asikainen et al.* [2012]'s corresponding results are shown in solid lines. The standard deviation (Figures 8a and 8c) and MAD (Figures 8b and 8d) are shown as vertical error bars in all $\alpha$ points.

### 6.1. Comparison With Earlier Studies

The first effort to quantify the $\alpha$ factors was done by *Asikainen and Mursula* [2011]. They identified cases where one old and one new satellites were close in time and space and assumed that the two satellites were observing the same particle population. The integral energy spectra from the two satellites were compared and the $\alpha$ factors for the old satellite were established. A challenge for the method is the lack of cases with new and old satellites being close in both time and space. Due to few cases, they did not derive $\alpha$ factors for the higher-energy channels. Different from our method is also the temporal evolution. *Asikainen and Mursula* [2011] fitted not only linear curves to the derived $\alpha$ factors but also second-, third-order polynomials, and PCHIP and presented $\alpha_{year}$ factors from the curve fitting. The follow-up paper, *Asikainen et al.* [2012], made use of the accumulated $Ap$ index to refine the estimate for the temporal evolution of the yearly $\alpha$ factor.

*Ødegaard* [2013] is a prestudy for our work here as it sorted the first year of SEM-2 MEPED proton data according to the $Kp$ index in a MLT/ILAT grid. The statistical maps gave the average proton flux as a function of MLT and ILAT. *Ødegaard* [2013] then compared measurements from the old satellite with a new satellite using these statistical maps. In that way, *Ødegaard* [2013] obtained $\alpha$ factors for P1 and P2. Like *Asikainen and Mursula* [2011] and *Asikainen et al.* [2012], this method also struggled regarding determination of the higher-energy $\alpha$ factors due to few cases.

Our study, as well as *Ødegaard* [2013] use average measurements to construct energy spectra for comparisons and mean or median to find the final $\alpha$ factor. But *Asikainen and Mursula* [2011] calculated numerous $\alpha$ factors for a limited number of conjunctions, whereas the final $\alpha$ factor was median of these.

Our study differs from *Asikainen and Mursula* [2011], *Asikainen et al.* [2012], and *Ødegaard* [2013] with a larger statistical database that makes us able to derive the $\alpha_{cal}$ factor also for the highest-energy channels of the proton MEPED detector. In addition, we make use of the newest satellite in orbit (MetOp-01) to find $\alpha_{cal}$ factors for the year 2013, and thereby $\alpha_{year}$ factors for the years 2011–2013.

In Figure 8 we focus on the NOAA 17 $\alpha_{cal}$ factors from a year that both *Asikainen and Mursula* [2011] and *Asikainen et al.* [2012] and our current study cover. Figure 8 shows $\alpha_{cal}$ factors as a function of nominal energy for our study (Figures 8a and 8c), *Asikainen and Mursula* [2011] and *Asikainen et al.* [2012] (Figures 8b and 8d). Solid lines with dark color visualize $\alpha_{cal}$ factors from 2007, while dotted lines with lighter colors in Figures 8a and 8c visualize $\alpha_{cal}$ factors from 2013. Figures 8b and 8d show the $\alpha_{cal}$ factors both before (dotted lines) and after this curve-fitting routine (dashed lines). Some of the $\alpha$ factors from *Asikainen and Mursula* [2011] are adjusted by *Asikainen et al.* [2012] due to the *Ap* activity index in order to get a better agreement between the satellites. The $\alpha$ factors resulting from the curve fit of *Asikainen et al.* [2012] are shown in solid lines in Figures 8b and 8d and are at times significantly different from the $\alpha$ factors achieved directly from their method shown as dotted and dashed lines. Our $\alpha$ factors are not adjusted due to curve fitting, and only one set of $\alpha$ factors are therefore shown in Figures 8a and 8c).

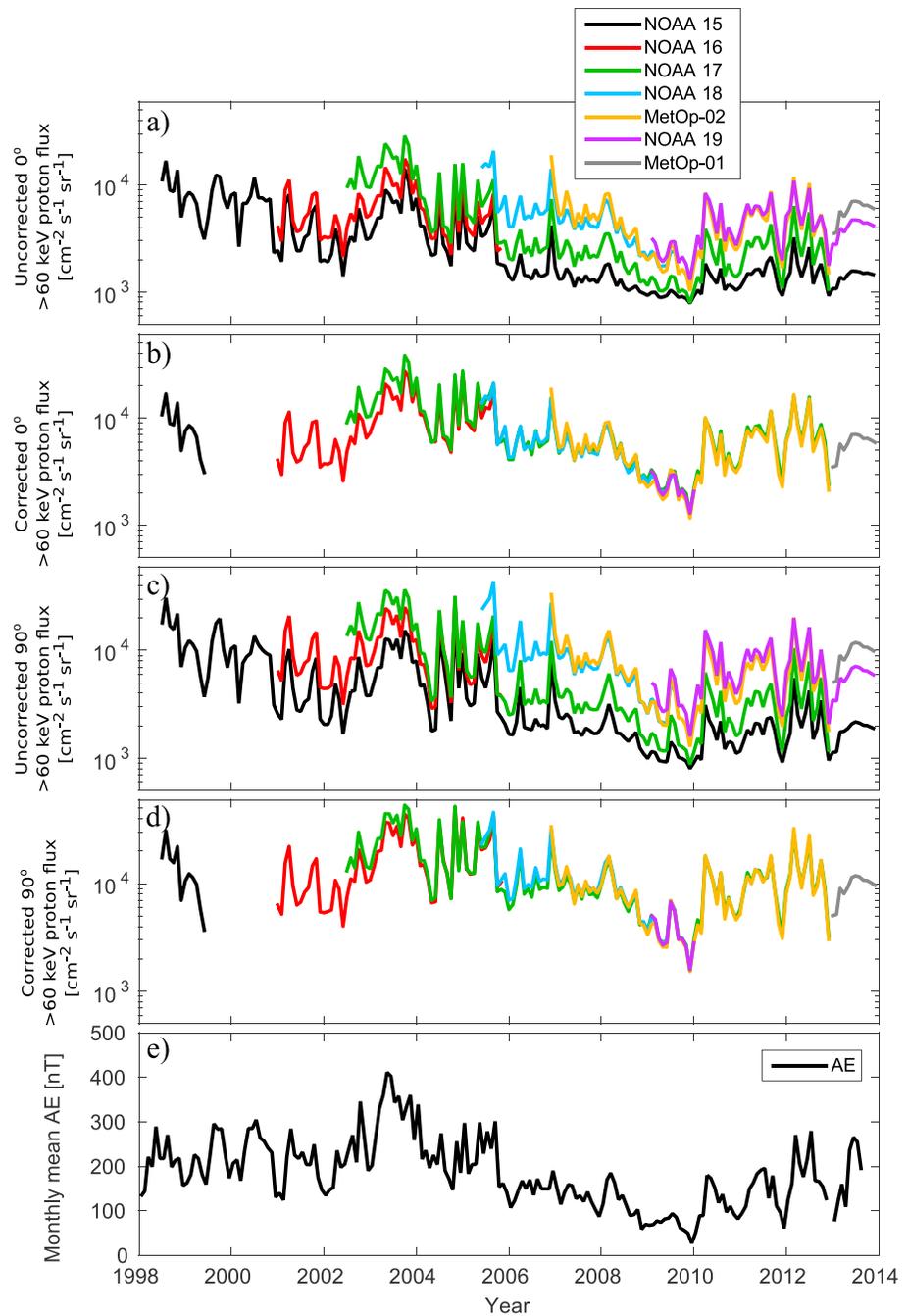

**Figure 9.** The monthly mean accumulated proton flux >60 keV measured by NOAA 15 (black line), NOAA 16 (red line), NOAA 17 (green line), NOAA 18 (light blue line), MetOp-02 (yellow line), NOAA 19 (purple line), and MetOp-01 (grey line). (a and b) Data from the 0° detector. (c and d) Data from the 90° detector. The uncorrected flux is shown in Figures 9a and 9c, and the alpha-corrected flux is shown in Figures 9b and 9d. Data from NOAA 15, NOAA 19 and MetOp-01 are added to the alpha-corrected (Figures 9b and 9d) but only for the first year of the satellite's operation. Data from NOAA 16 and NOAA 18 are displayed in the alpha-corrected panels (Figures 9b and 9d) until 1 year beyond their last $\alpha$ factor. (e) The monthly mean *AE* index.

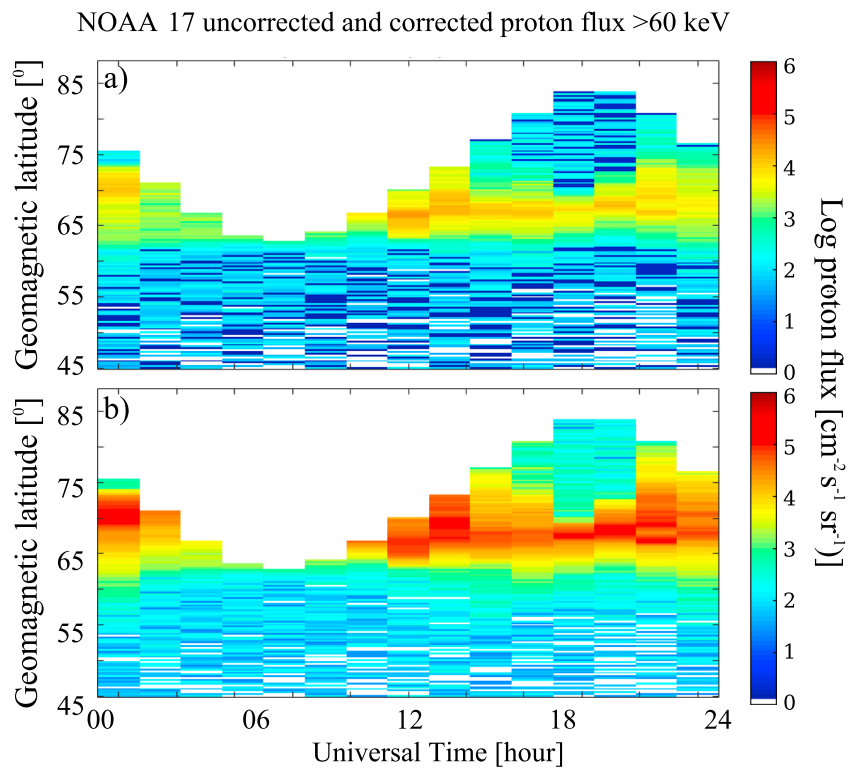

**Figure 10.** (a) Uncorrected and (b) corrected proton flux > 60 keV on a logarithmic color scale. The NOAA 17 proton flux are from the 90° detector. The data are from the evening side Northern Hemisphere on 5 January 2012.

Comparing our $\alpha_{cal}$ factors from NOAA 16 in 2005, NOAA 17 in 2007, and NOAA 18 in 2009 with *Asikainen et al.* [2012], we find that their new energy thresholds show about 10% discrepancies from ours for the two lowest-energy channels. Their energy threshold for channel P3 is generally elevated by 5%–40% compared to ours.

Figure 8 shows that P2 degrades faster than P1, as visualized by the $\alpha_1$ and $\alpha_2$ factors. The degradation factors of all satellites examined in this study reveal the same quite puzzling feature regarding the temporal evolution between the $\alpha_1$ and $\alpha_2$ factors. The studies by *Asikainen and Mursula* [2011], *Asikainen et al.* [2012], and *Ødegaard* [2013] (not shown) all display the same trend, implying that this is a real feature of the detector degradation. Although a physical understanding of this artifact would be interesting, it is not essential for establishing a set of $\alpha$ factors that can be applied to the NOAA POES data.

## 7. Application of the $\alpha$ Factors

The importance of having high and reliable accuracy in the $\alpha$ factors is best illustrated by an example. The $\alpha$ factors are given as a set $[\alpha_{P1}, \alpha_{P2}, \alpha_{P3}, \alpha_{P4}, \alpha_{P5}]$ as shown in equation (2). For each of the nominal energies given in Table 1, one has to multiply by the $\alpha$ factor to achieve the corrected energy threshold:

$$E_{\text{corrected}}(n) = E_{\text{nominal}}(n) \times \alpha(Pn) \tag{10}$$

where $n \in [1, 2, 3, 4, 5]$ is the energy channel. The shape of the spectrum is essential for the level of correction at a specific energy.

Figures 9a and 9c show the uncorrected monthly mean integral fluxes for all satellites interpolated to >60 keV. This is done in order to compare at the same energy ranges, as 60 keV corresponds to the most degraded satellite in the period considered. The fluxes for all satellites in Figures 9a and 9c exhibit roughly the same short-term variations, but the measured flux is significantly lower when measured by an old satellite compared with a new one. Figures 9b and 9d show the mean integral fluxes for each month and satellite after the fluxes are corrected using the corresponding $\alpha_{month}$ factor. There is now a close to perfect overlap between the fluxes measured by the different satellites. Over the years 1998 to 2013 it is evident that there is a steady decline in the radiation measured at low altitude by a factor of 2 roughly linked to the solar cycle shown in Figure 1. The short time variation in the data is due to geophysical activity with injection of protons into the magnetosphere, as evident from the close similarities between these variations and the monthly mean *AE* index shown in Figure 9e.

Finally, to further illustrate the importance of applying the corrections, Figure 10 shows data from the 90° detector on board NOAA 17, from the evening side Northern Hemisphere on 5 January 2012, after the detector has experienced 10 years of radiation and subsequent degradation. For each of the pixels, the integral flux from all energy channels are used to construct a PCHIP-fitted spectrum. The spectrum is then corrected using the specific $\alpha_{month}$ factors. The uncorrected and the corrected spectra are used to determine the integral flux of $E > 60$ keV shown in Figures 10a and 10b, respectively. The corrected flux is often nearly 1 order of magnitude larger than the uncorrected flux.

## 8. Conclusion

For the first time the level of degradation of the MEPED proton detectors is revealed for all five energy channels. Our $\alpha_{cal}$ factors are based on a larger database than earlier studies. We also separate the satellites according to MLT sectors, and the new MetOp-01 satellite gives us new $\alpha_{cal}$ factors for 2013 for the satellites in the same MLT sector. Figure 9 shows that the corrected fluxes overlap close to perfectly, verifying the quality of the derived correction factors.

It is important in quantitative studies to correct for changes in energy levels in the proton detector as the detectors experience radiation damage throughout their lifetime. In the future, there will be no new satellite to compare against. It would therefore be valuable to establish a method to predict the degradation in order to have a continuous set of correction factors throughout the lifetime of all NOAA and MetOp satellites.

**Acknowledgments**

This study was supported by the Research Council of Norway under contracts 184701, 212014, and 223252. The authors especially thank David S. Evans for all help and comments. The authors thank Janet C. Green, Daniel C. Wilkinson, and the NOAA's National Geophysical Data Center (NGDS) for providing NOAA data. We thank the SIDC-team, World Data Center for the Sunspot Index, Royal Observatory of Belgium, Monthly Report on the International Sunspot Number, online catalog of the sunspot index: http://www.sidc.be/sunspot-data/. We thank NOAA and European Space Agency (ESA) for providing information regarding ascending node of the satellites (http://www.ngdc.noaa.gov/ and http://www.esa.int/esaLP/).